\documentclass[twocolumn,showpacs,prb,floatfix,superscriptaddress,citeautoscript,cite]{revtex4}
\usepackage{graphicx}
\usepackage{color}
\usepackage{amsmath}

\begin{document}

\title{$h$-AlN-Mg(OH)$_{2}$ vdW Bilayer Heterostructure: Tuning the excitonic characteristics}

\author{C. Bacaksiz}
\affiliation{Department of Physics, Izmir Institute of Technology, 35430, Izmir, Turkey}

\author{A. Dominguez}
\affiliation{Max Planck Institute for the Structure and Dynamics of Matter, Hamburg, Germany}

\author{A. Rubio}
\affiliation{Nano-Bio Spectroscopy Group and ETSF, Dpto. Fisica de Materiales, 
Universidad del Pais Vasco, CFM CSIC-UPV/EHU-MPC and DIPC, 20018 San Sebastian, 
Spain}
\affiliation{Max Planck Institute for the Structure and Dynamics of Matter and 
Center for Free-Electron Laser Science, Luruper Chaussee 149, 22761 Hamburg, 
Germany}

\author{R. T. Senger}
\affiliation{Department of Physics, Izmir Institute of Technology, 35430, Izmir, Turkey}


\author{H. Sahin}
\affiliation{Department of Photonics, Izmir Institute of Technology, 35430, Izmir, Turkey}


\date{\today}

\begin{abstract}

Motivated by recent studies that reported the successful synthesis of monolayer Mg(OH)$_{2}$ [Suslu \textit{et al.}, 
Sci. Rep. \textbf{6}, 20525 (2016)] and hexagonal (\textit{h}-)AlN [Tsipas \textit{et al}., Appl. Phys. Lett. 
\textbf{103}, 251605 (2013)], we investigate structural, electronic, and optical properties of vertically stacked 
$h$-AlN and Mg(OH)$_{2}$, through \textit{ab initio} density-functional theory (DFT), many-body quasi-particle 
calculations within the GW approximation, and the Bethe-Salpeter equation (BSE). It is obtained that the bilayer 
heterostructure prefers the $AB^{\prime}$ stacking having direct band gap at the $\Gamma$ with Type-II band alignment 
in 
which the valance band maximum and conduction band minimum originate from different layer. Regarding the optical 
properties, the imaginary part of the dielectric function of the individual layers and hetero-bilayer are investigated. 
The hetero-bilayer possesses excitonic peaks which appear only after the construction of the hetero-bilayer. The lowest 
three exciton peaks are detailedly analyzed by means of 
band decomposed charge density and the oscillator strength. Furthermore, the wave function calculation shows that the 
first peak of the hetero-bilayer originates from spatially indirect exciton where the electron and hole localized at 
$h$-AlN and Mg(OH)$_{2}$, respectively, which is important for the light harvesting applications. 
\end{abstract}

\maketitle

\section{Introduction}

After the discovery of graphene\cite{Novoselov2}, interest in atomically thin materials\cite{Novoselov1} has grown 
rapidly due to their extraordinary physical properties.\cite{Butler,Chhowalla} In the last decade, several 2D materials 
have been synthesized and theoretically predicted, such as silicene,\cite{Cahangirov,Kara} 
germanene,\cite{Cahangirov} stanene,\cite{Guzman,Bechstedt} transition metal dichalcogenides (TMDs such 
as MoS$_{2}$, WS$_{2}$),\cite{Gordon,Coleman,Wang1,Ross,Sahin2,Tongay,Horzum,Chen3} and III-V binary compounds (e.g. 
$h$-BN, $h$-AlN).\cite{Sahin3,Wang2,Kim,Tsipas,Bacaksiz} Beside the single crystal of 2D materials, recently emerging 
field is their vertically stacked heterostructures.\cite{Geim} Because of the van der Waals type weak interlayer 
interaction, the synthesis of heterostructures is not restricted with the lattice matching of the each layers. This 
provides a wide variety of combinations of layers which exhibit different electronic and optical 
properties.\cite{Britnell,Fang,Lee2,Hunt,Hong,Ferrari,Tan}

Stable hexagonal crystalline structure of AlN was first theoretically predicted by Sahin \textit{et al}.\cite{Sahin3} 
and experimentally synthesized by Tsipas \textit{et al.}\cite{Tsipas} They found that, differing from its bulk 
structure which is an insulator, monolayer $h$-AlN is a semiconductor with the indirect band gap where the valance band 
maximum and conduction band minimum at the K and $\Gamma$ points, respectively. Almeida \textit{et al.}\cite{Almeida} 
investigated properties of defects, such as vacancies, antisites, and impurities, in $h$-AlN. It was  reported that N 
vacancies and Si impurities lead to the breaking of the planar symmetry and cause significant changes in the electronic 
properties. Shi \textit{et al.}\cite{Shi} calculated the magnetic properties of bare and transition-metal (TM) doped 
AlN nanosheets by using first-principles calculations. They reported that nonmagnetic $h$-AlN can be magnetized upon a 
single TM atom. Moreover, the electronic structures of nanoribbon of AlN were investigated by Zheng \textit{et 
al.}\cite{Zheng} and it is predicted that zig-zag edge nanoribbons have an indirect band gap while armchair edge 
nanoribbons have a direct band gap, and band gaps of both zig-zag and armchair decrease monotonically with increasing 
ribbon width. Furthermore, Bacaksiz \textit{et al.}\cite{Bacaksiz} reported that the electronic band 
structure changes significantly when the number of layer increases and the structure having 10 or more layers exhibits 
direct bandgap as bulk form. More recently, Kecik \textit{et al.}\cite{Kecik} investigated the optical properties 
of mono- and few-layer $h$-AlN under strain. They reported that the absorption peaks stand outside the visible-light 
regime, on the other hand, the applied tensile strain gradually redshifts the optical spectra.

As a constituent of hetero-bilayer, physical properties of Mg(OH)$_{2}$ were investigated previously in several 
studies.\cite{Utamapanya,Ding,Sideris} It is a layered metal hydroxide with a wide band gap\cite{Murakami}. As a member 
of the alkaline-earth-metal hydroxides (AEMHs) family, Ca(OH)$_{2}$ was predicted to be stable in the bilayer and 
monoayer forms by Aierken \textit{et al.}\cite{Aierken} In addition, Torun \textit{et al.} investigated the optical 
properties of GaS-Ca(OH)$_{2}$ bilayer heterostructure by using GW+BSE and reported that in spite of the similarities 
of electronic structures, the different stacking types have different optical spectra.\cite{Torun2} Recently, Tsukanov 
\textit{et al.} investigated the interaction of organic anions with layered double hydroxide nanosheets consist of 
Mg and Al by using molecular dynamics simulations.\cite{Tsukanov} In addition, very recently, the monolayer 
Mg(OH)$_{2}$ was synthesized and reported that in spite of the optically inactive nature of Mg(OH)$_{2}$, the 
photoluminescence intensity of monolayer MoS$_{2}$ was assisted by Mg(OH)$_{2}$ and enhanced\cite{Suslu}. Most 
recently, Yagmurcukardes \textit{et al.} investigated the hetero-bilayer of the Mg(OH)$_{2}$ and 
WS$_{2}$.\cite{Yagmurcukardes} They reported that the lower energy optical spectrum of the Mg(OH)$_{2}$-WS$_{2}$ 
hetero-bilayer is dominated by the excitons originates WS$_{2}$ layers.  These two studies 
indicate that the Mg(OH)$_{2}$ is a candidate for tuning the optical property of other monolayer materials.

In the present work, the heterostructure of two perfectly-matching monolayers of $h$-AlN, a member of III-V binary 
compounds, and Mg(OH)$_{2}$, a member of alkaline-earth-metal hydroxide (AEMH), are considered. The similar lattice 
constants of the layers provide us to use smaller supercell which is important especially for the calculating the 
optical properties. We found that the 
vertically stacked heterostructure possesses excitonic peaks which appear only after the construction of the 
hetero-bilayer. More significantly, the wave function 
calculation shows that the first peak of the imaginary part of the dielectric function for the hetero-bilayer 
originates from spatially indirect exciton where the 
electron and hole localized at $h$-AlN and Mg(OH)$_{2}$, respectively.

The paper is organized as follows: In Sec. \ref{comp} we give details of our computational methodology. In Sec. 
\ref{monolayers} we present a brief overview of the structural and electronic properties of monolayer $h$-AlN and 
Mg(OH)$_{2}$. In Sec. \ref{bilayer_hetero} different stacking orders of bilayer heterostructure of $h$-AlN and 
Mg(OH)$_{2}$ and also optical property of the energetically favorable stacking order are investigated in detail. 
Finally, we present our conclusion in Sec. \ref{conc}.

\section{Computational Methodology}\label{comp}

Employing (DFT)-based methods, we investigated the the structural, electronic and optical properties of monolayer 
$h$-AlN and Mg(OH)$_{2}$ and their hetero-bilayers. We used the Vienna ab-initio simulation package 
VASP\cite{vasp1,vasp2,vasp3} which solves the Kohn-Sham equations iteratively using a plane-wave basis set. To describe 
electron exchange and correlation, the Perdew-Burke-Ernzerhof (PBE) form of the generalized gradient approximation 
(GGA)\cite{GGA-PBE} was adopted. The van der Waals (vdW) forces which is important for layered materials was taken into 
account by using the DFT-D2 method of Grimme.\cite{vdW1,vdW2} To obtain partial charge on the atoms, a Bader charge 
analysis was used.\cite{Bader1}

Structural optimizations were performed with the following parameters. The kinetic energy cut-off of the plane-wave 
basis set was 500 eV in all calculations. The total energy difference between the sequential steps in the iterations 
was taken 10$^{-5}$ units as convergence criterion. The convergence for the Hellmann-Feynman forces per unit cell atom 
was taken to be 10$^{-4}$ eV/\AA{}. Gaussian smearing of 0.05 eV was used and the pressures on the unit cell were 
decreased to a value less then 1.0 kB in all three directions. For the determination of accurate charge densities, 
Brillouin zone integration was performed using a $35\times35\times1$ $\Gamma$-centered mesh for the primitive unit 
cell. To avoid interactions between adjacent monolayer and hetero-bilayers, our calculations were performed with a 
large unit cell including $\sim$18 \AA{} vacuum space. We also calculated the cohesive energy ($E_{coh}$), which was 
formulated as $ E_{coh} =   [ ( \sum\limits_{i}  n_{i} E_{i} )  - E_{T} ]  / \sum\limits_{i} n_{i} $ where $i$ stands 
for the atoms which contract the structure; $n_{i}$, $E_{i}$, and $E_{T} $ are the number of $i$ atom in the unit cell, 
the energy of free $i$ atom, and the total energy per unit cell. 

\begin{figure}[htbp]
\includegraphics[width=8.5 cm]{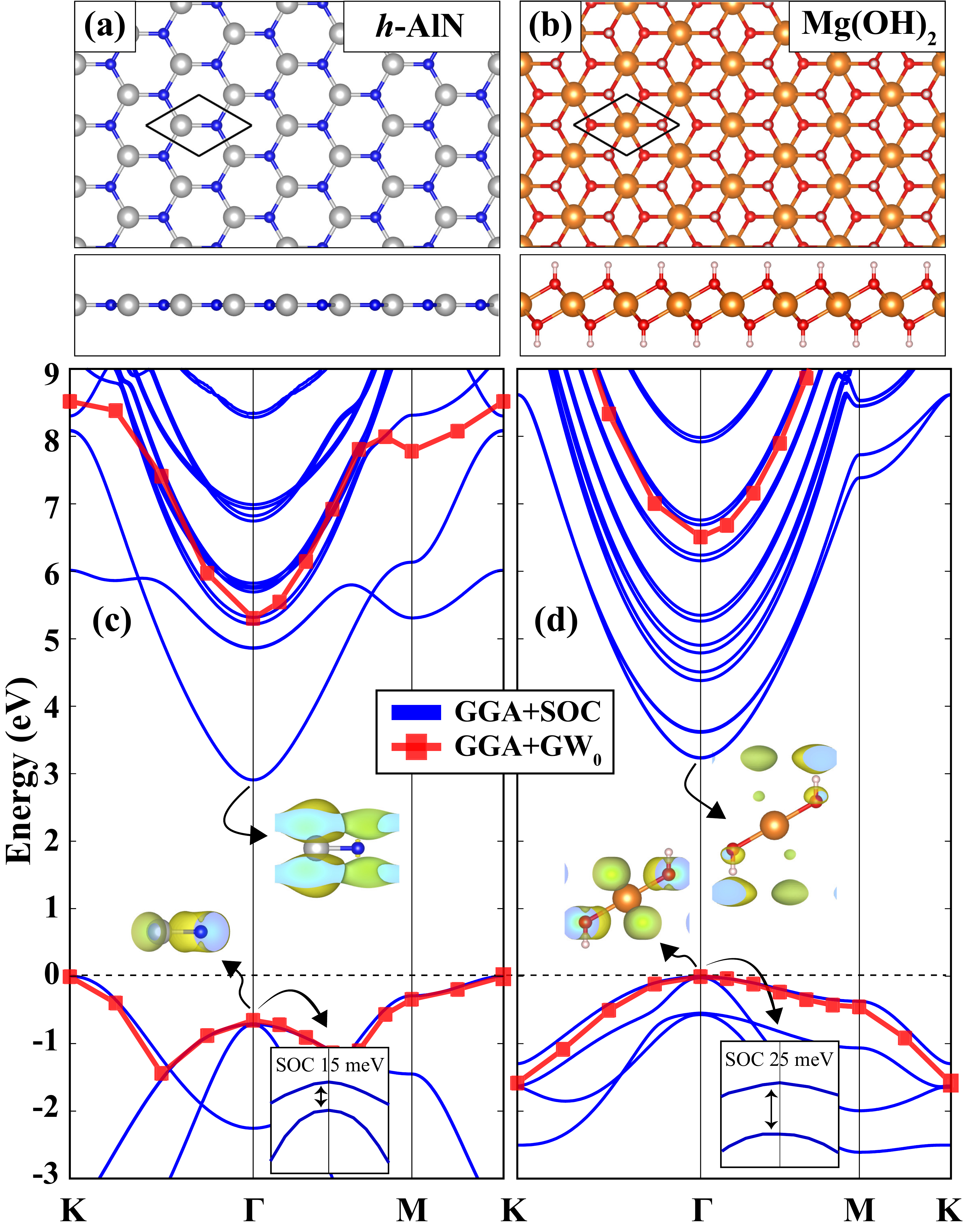}
\caption{\label{1lfig}
(Color online) Upper panel illustrates the structure of monolayer $h$-AlN (left) and Mg(OH)$_{2}$ (right). (a) and (b) 
are the structures (top and side view), the solid black parallelograms show the unitcell of the structural, (c) and (d) 
are the band structures of $h$-AlN and Mg(OH)$_{2}$, respectively. The blue and light-red curves with square are for 
GGA+SOC and GGA+GW$_{0}$, respectively. The dashed vertical line is shows the Fermi energy. In the band diagram, the 
charge densities of the VBM and CBM and the spin-orbit splitting at the $\Gamma$ point.   }
\end{figure}

\begin{table*}[htbp]
\caption{\label{1ltable} Calculated parameters for monolayer $h$-AlN and Mg(OH)$_{2}$ are the lattice constant in the 
lateral direction, $a$; the final charges of Al, N, Mg, O, and H, $\rho_{\text{Al}}$, $\rho_{\text{N}}$, 
$\rho_{\text{Mg}}$,  $\rho_{\text{O}}$, $\rho_{\text{H}}$, respectively; the work function $\Phi$; the cohesive energy, 
E$_{coh}$; and the spin-orbit splitting, $\Delta_{SO}$. E$_{g}^{\text{GGA}}$ and E$_{g}^{\text{GW$_{0}$}}$ are the 
energy band gap values within GGA+SOC and GGA+GW$_{0}$, respectively.}
\begin{tabular}{lccccccccccccccc}
\hline\hline
   &a    &$\rho_{\text{Al}}$&$\rho_{\text{N}}$&$\rho_{\text{Mg}}$&$\rho_{\text{O}}$&$\rho_{\text{H}}$&$\Phi$& 
E$_{coh}$&E$_{g}^{\text{GGA}}$ &$\Delta_{SO}$&$\Gamma\longrightarrow\Gamma$&$K\longrightarrow 
K$&E$_{g}^{\text{GW$_{0}$}}$\\
&(\AA)&($e^{-}$)        & ($e^{-}$)      &($e^{-}$)        &($e^{-}$)       &($e^{-}$) &(eV)  & (eV)   & (eV)  &(meV)& 
  (eV)& (eV)&(eV)            \\
\hline
$h$-AlN     & 3.13 &0.7 &     7.3           &      -        &        -        &  -      &5.12  & 5.35     & 2.90 
     
  &15& 3.62 & 6.02& 5.37 ($K-\Gamma$)    \\
Mg(OH)$_{2}$& 3.13 &  -  &         -          &    0.3       &    7.4         & 0.4   &4.20  & 4.38     & 3.23 
    
  &25& 3.23 & 8.92& 6.51 ($\Gamma-\Gamma$)    \\
\hline\hline 
\end{tabular}
\end{table*}

In addition, the quasi-particle (QP) energies were calculated within the GW$_{0}$ approximation where the 
single-particle Green function part (G) was iterated and the screened Coulomb interaction part (W) was fixed. On top of 
GW$_{0}$ approximation, the energies of two particle system of quasi-electron and quasi-hole (exciton) were calculated 
by solving BSE\cite{bse,Hanke}. To obtain accurate QP states, 320 bands were considered in the GW$_{0}$ calculations. 
Our convergence tests showed that the $12\times12\times1$ k-point sampling and the vacuum spacing of $\sim$28 \AA{} 
well approximates the excitonic properties of vdw hetero-bilayer structure (See Appendix). The 8 highest valence 
bands and lowest conduction bands were considered as a basis for the excitonic states.

\begin{figure}[htbp]
\includegraphics[width=8.5 cm]{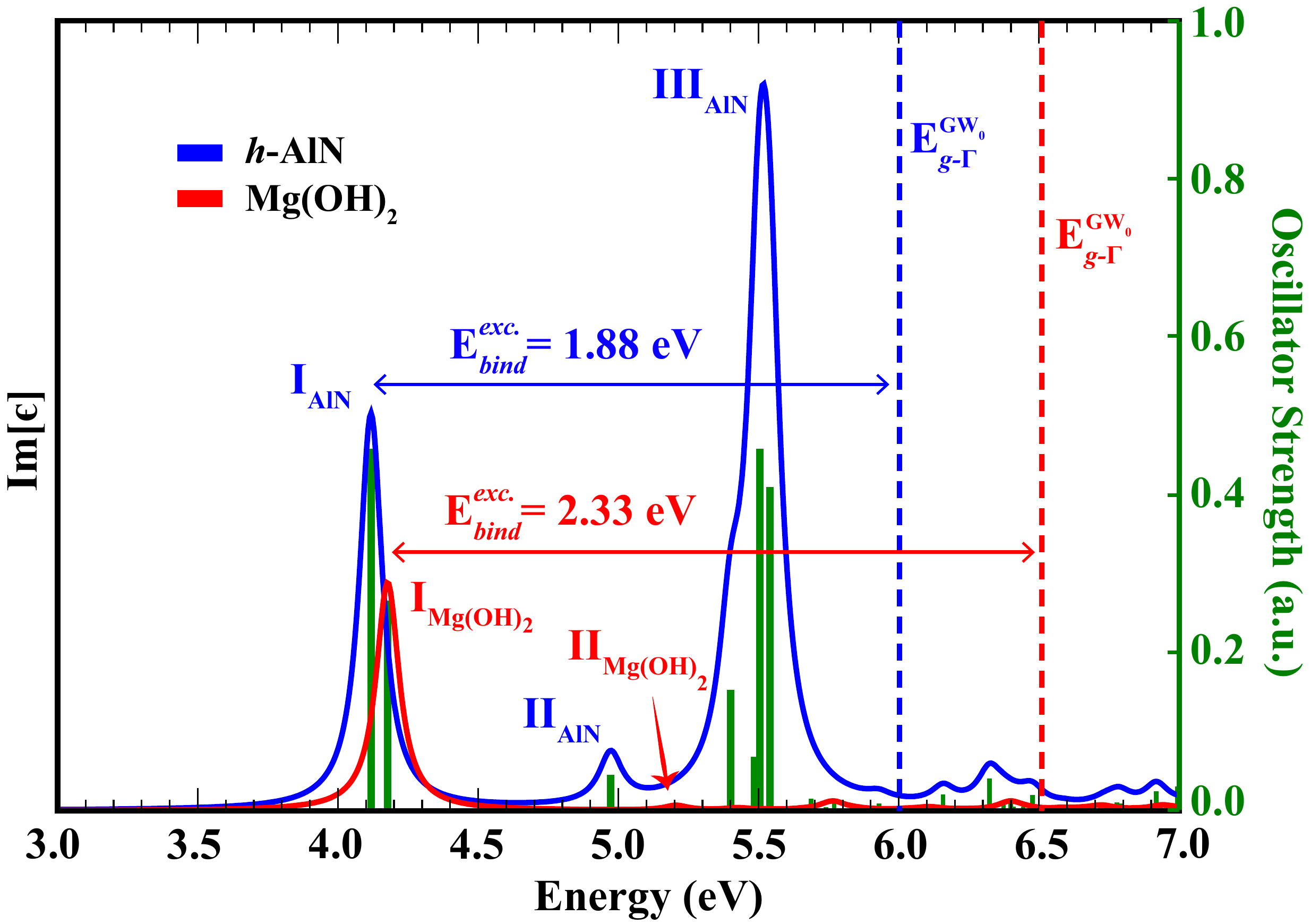}
\caption{\label{bse_single}
(Color online) The imaginary part of the dielectric function of $h$-AlN (blue curve) and Mg(OH)$_{2}$ (red curve) with 
oscillator strength of excitonic states (vertical green lines). The first, second and third excitons are labeled as I, 
II, III with the subscript of the corresponding individual layer, respectively. The quasi-particle band gap of $h$-AlN 
(blue dashed) and Mg(OH)$_{2}$ (red dashed) at $\Gamma$ and exciton binding energies (E$_{bind}^{exc.}$) are given. }
\end{figure}

\section{Single Layer $h$-A\lowercase{l}N \lowercase{and} M\lowercase{g}(OH)$_{2}$}\label{monolayers}

Before analysis of bilayer heterostructure of $h$-AlN and Mg(OH)$_{2}$, the monolayer constituents are discussed. 
Firstly, the monolayer structure of $h$-AlN (see Fig.~\ref{1lfig}(a)) belongs to space group $P6_{3}/mmc$ with the 
lattice constant 3.13 \AA{} which is consistent with that of few-layer hexagonal AlN that reported by Tsipas 
and the co-workers.\cite{Tsipas} In Table~\ref{1ltable}, the number of valance electrons of the isolated single 
atom and of the one in crystal are given. It is seen that Al donates 2.3 of 3.0 $e^{-}$ to N which indicates that the 
bond between Al and N have strong ionic character. The work function is found to be 5.12 eV. As shown in 
Fig.~\ref{1lfig}(c), $h$-AlN has an indirect band gap of 2.9 eV where the valance band maximum (VBM) and the conduction 
band minimum (CBM) are at the $\Gamma$ and $K$ points, respectively. GW$_{0}$ band gap is calculated to be 5.30 eV. In 
addition, differing from similar TMDs, the spin-orbit (SO) splitting at the $\Gamma$ point is quite 
small (15 meV).\cite{Mak} 

Structural and electronic properties of monolayer Mg(OH)$_{2}$ which is another building-block of hetero-bilayer are 
also presented in Figs.~\ref{1lfig}(b) and (d). The lattice parameter  of Mg(OH) which belongs to $P\bar{3}m1$ space 
group is found to be 3.13 \AA{}. Therefore, it perfectly matches to $h$-AlN. Upon the formation of this crystalline 
structure In this structure: (i) Mg donates almost all valance electrons, 1.8 of 2.0 $e^{-}$, and (ii) H donates 0.6 of 
1.0 $e^{-}$ to O atom. The work function is calculated to be 4.20 eV. Mg(OH)$_{2}$ has direct band gap of 3.23 eV, 6.51 
within GGA and GGA+GW$_{0}$, respectively. The wide GW$_{0}$ band gap is close to the value which is previously 
obtained.\cite{Suslu} The SO splitting at the $\Gamma$ is 25 meV which is slightly larger than that of $h$-AlN (15 
meV). This small difference can be understood from the difference between the atomic radius of N and O atoms.

In Fig.~\ref{bse_single}, the energy dependency of the imaginary part of the dielectric function of pristine $h$-AlN 
and Mg(OH)$_{2}$ are shown. We name the peaks as I, II, and III (first, second and third, respectively) with the 
subscript that specify the corresponding individual layer. The I$_{AlN}$ appears at 4.12 eV with the exciton binding 
energy of 1.88 eV. The II$_{AlN}$ and III$_{AlN}$ appear at 4.97 and 5.4 eV, respectively. The III$_{AlN}$ consists of 
4 exciton levels. The I$_{Mg(OH)_{2}}$ and II$_{Mg(OH)_{2}}$ appear at 4.18 and 5.20 eV, respectively. The excitonic 
binding energy of the I$_{Mg(OH)_{2}}$ is found to be 2.33 eV. The peaks of the $h$-AlN are larger than those of 
Mg(OH)$_{2}$. The reason is that all the peaks shown in Fig.~\ref{bse_single} originate from the excitation at the 
$\Gamma$, hence the overlapping of electron and hole states are larger for the $h$-AlN which is clearly seen at the 
charge density of band edges in Fig.~\ref{1lfig}. In addition, the excitonic binding energies are larger compared to 
the bulk materials. Such binding energies were previously reported and discussed.\cite{Wirtz,Pishtshev,Cudazzo}

\section{Bilayer Heterostructure}\label{bilayer_hetero}

In this section, we give an analysis of structural, electronic and optical properties of bilayer heterostructure of 
vertically stacked monolayer $h$-AlN and Mg(OH)$_{2}$.

\begin{figure}[htbp]
\includegraphics[width=8.5 cm]{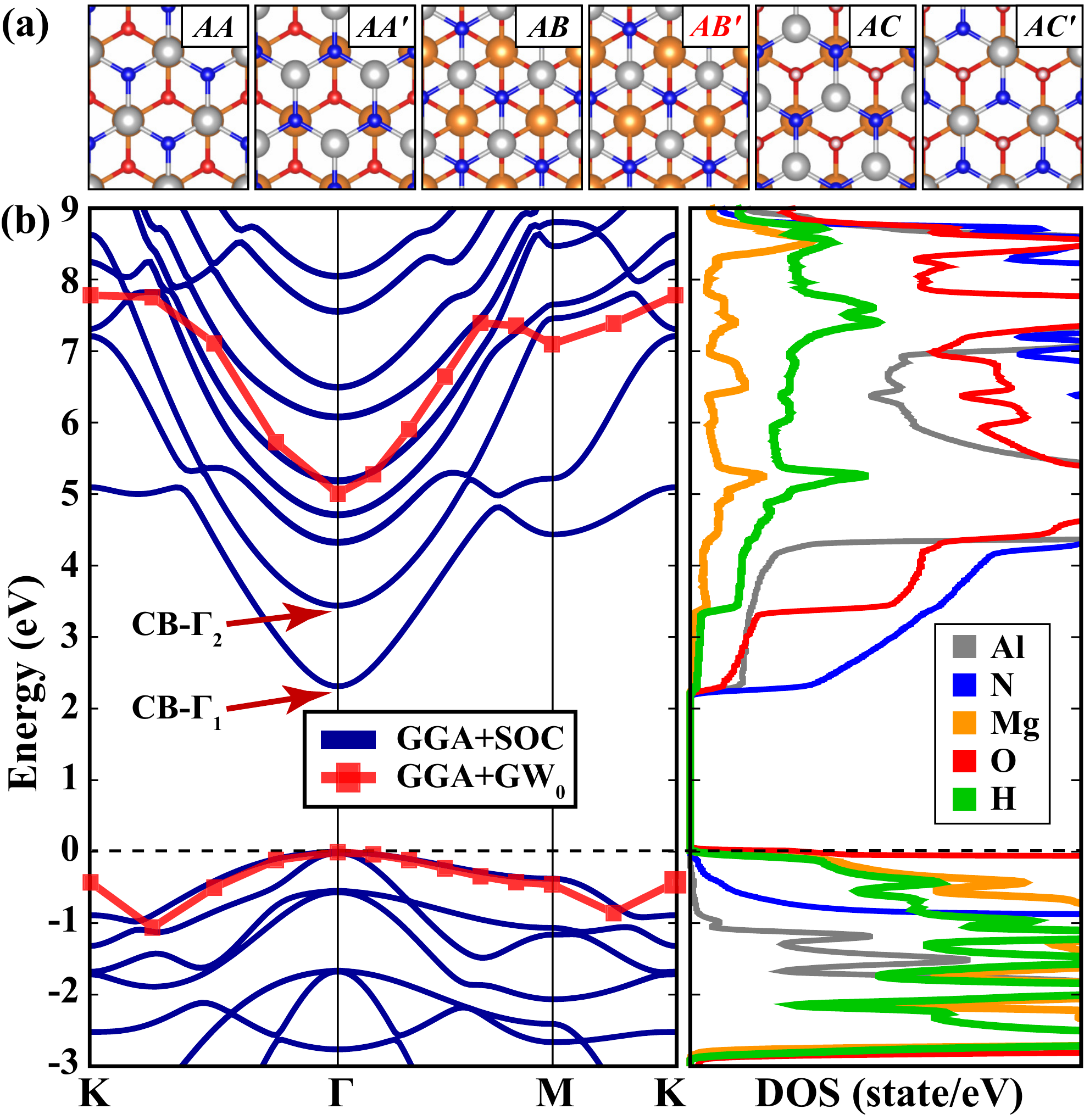}
\caption{\label{hetero_fig}
(Color online) (a) Top view of six different stacking orders, (b) Electronic band dispersion (left panel) and the 
partial density of state (right panel) of $AB^{\prime}$ stacking order which is the most favorable stacking order. The 
CB-$\Gamma_{1}$ and  CB-$\Gamma_{2}$ label the first and second lowest levels of the CB at $\Gamma$ point, 
respectively.}
\end{figure}

\begin{table}[htbp]
\caption{\label{hetero_table} The stacking labels are given in the first column. Calculated parameters for bilayer 
heterostructure of $h$-AlN and Mg(OH)$_{2}$ are; the interlayer distance, $d_{L-L}$; the energy difference between the 
ground state, $\Delta_{E}$; the interaction energy between the layers, E$_{int}$; and the work function $\Phi$. 
E$_{g}^{\text{GGA}}$  are the energy band gap values within  GGA+SOC.}
\begin{tabular}{lcccccccc}
\hline\hline
             &$d_{L-L}$& $\Delta_{E} $&  E$_{int}$ &$\Phi$& E$_{g}^{\text{GGA}}$ \\
             &(\AA{})& (meV)        &   (meV)      &(eV)  & (eV)                    \\
\hline
$AB^{\prime}$  &2.13&     0        &    237        &4.31   &  2.30               \\
$AA$           &2.12&     6        &    231        &4.32   & 2.31               \\
$AC$           &2.06&    66        &    171        &4.70   &  2.65              \\
$AC^{\prime}$  &2.10&    76        &    161        &4.69   & 2.61               \\
$AA^{\prime}$  &2.75&   162        &    75         &4.35   &  2.14              \\
$AB$           &2.75&   164        &    73         &4.29   &  2.14               \\
\hline\hline 
\end{tabular}
\end{table}

\subsection{Determination of Stacking Type}

First of all, to determine energetically favorable structure, six different stacking orders are considered. As shown in 
Fig.~\ref{hetero_fig}(a), possible stacking orders are: $AA$ (Al on Mg, N on H), $AA^{\prime}$ (N on Mg, Al on H), $AB$ 
(Al on H, N at the midpoint of three H), $AB^{\prime}$ (N on H, Al at the midpoint of three H), $AC$ (Al and N on 
top of midpoint of H triangles and N on Mg), and $AC^{\prime}$ (switched of Al and N form of $AC$). The energy 
difference between different stackings are given in Table~\ref{hetero_table}. In general, the hetero-bilayer has three 
type of properties with respect to localization of Al or N on trigonal H surface of Mg(OH)$_{2}$; when N atom is on top 
of H atom, which means the $AB^{\prime}$ and $AA$ stacking types, the hetero-bilayers posses the lowest energies with 
highest binding energies. They have also similar band gap of 2.30 and 2.31 eV, respectively. The work functions are 
found to be 4.31 and 4.32 eV which are very close to each other. Instead of these energetically favorable forms, the 
minimum interlayer distances are obtained when the Al and N atoms coincide to the midpoint H triangles ($AC^{\prime}$ 
and $AC^{\prime}$). The band gap and work function values of the $AC^{\prime}$ and $AC^{\prime}$ stacking are found to 
be similar as well. When Al is on top of H , the structures have the highest interlayer distances highest energy and 
lowest binding energy. The $AA$ with $AB^{\prime}$, $AA^{\prime}$ with $AB$, and $AC$ with $AC^{\prime}$ have similar 
parameters.

The minimum energy of the hetero-bilayer is obtained when the layer are stacked in the form of $AB^{\prime}$ stacking 
which exhibits the maximum interlayer interaction of 237 meV. Interlayer distance, which is defined as the 
perpendicular distance from surface H atom to the $h$-AlN plane, is 2.13 \AA{}. The work function of Mg(OH)$_{2}$ side 
is calculated to be 4.31 eV. The VBM and CBM are at $\Gamma$ as in monolayer Mg(OH)$_{2}$, however the band gap values 
of 2.30 eV within GGA which are significantly low as compared to that of monolayer of $h$-AlN and Mg(OH)$_{2}$. In 
addition, the $AA$ stacking order is energetically very close to $AB^{\prime}$. For both $AB^{\prime}$ and $AA$, the 
layer-layer interaction, interlayer distances, and also band gap values are almost same. 

In addition, considering vacuum level positions, formation of a staggered gap, with a mismatch of 1.02 eV at the 
intimate contact point of monolayers is predicted. Moreover, Since the minimum energy difference between the VB and CB is 
at $\Gamma$ point (see Fig.~\ref{hetero_fig} (b)), the lower energy optical activity takes place at this high symmetry 
point of Brillouin Zone. Such a contact between two atomically-flat surfaces forms an 
ultra-clean Type-II heterojunction.

\subsection{Origin of Excitonic States of Bilayer Heterostructure}

In this part, the origin of the prominent excitonic peaks in the AlN/Mg(OH)$_{2}$ hetero-bilayer structure, which 
corresponds to the AB$^{\prime}$  stacking, are investigated in details. There are two effects that determine 
characteristic properties of exciton in two-dimensional materials: the dielectric screening and the structural 
confinement. In general, when the crystal structure of a usual semiconductor is reduced from 3D to 2D, the dielectric 
screening between electron and hole takes place only inside layer. Therefore, as a consequence of the dimensional 
reduction, the screening effect on electron-hole pair is lower for 2D crystals as compared to the bulk.

\begin{figure}
\includegraphics[width=8.5 cm]{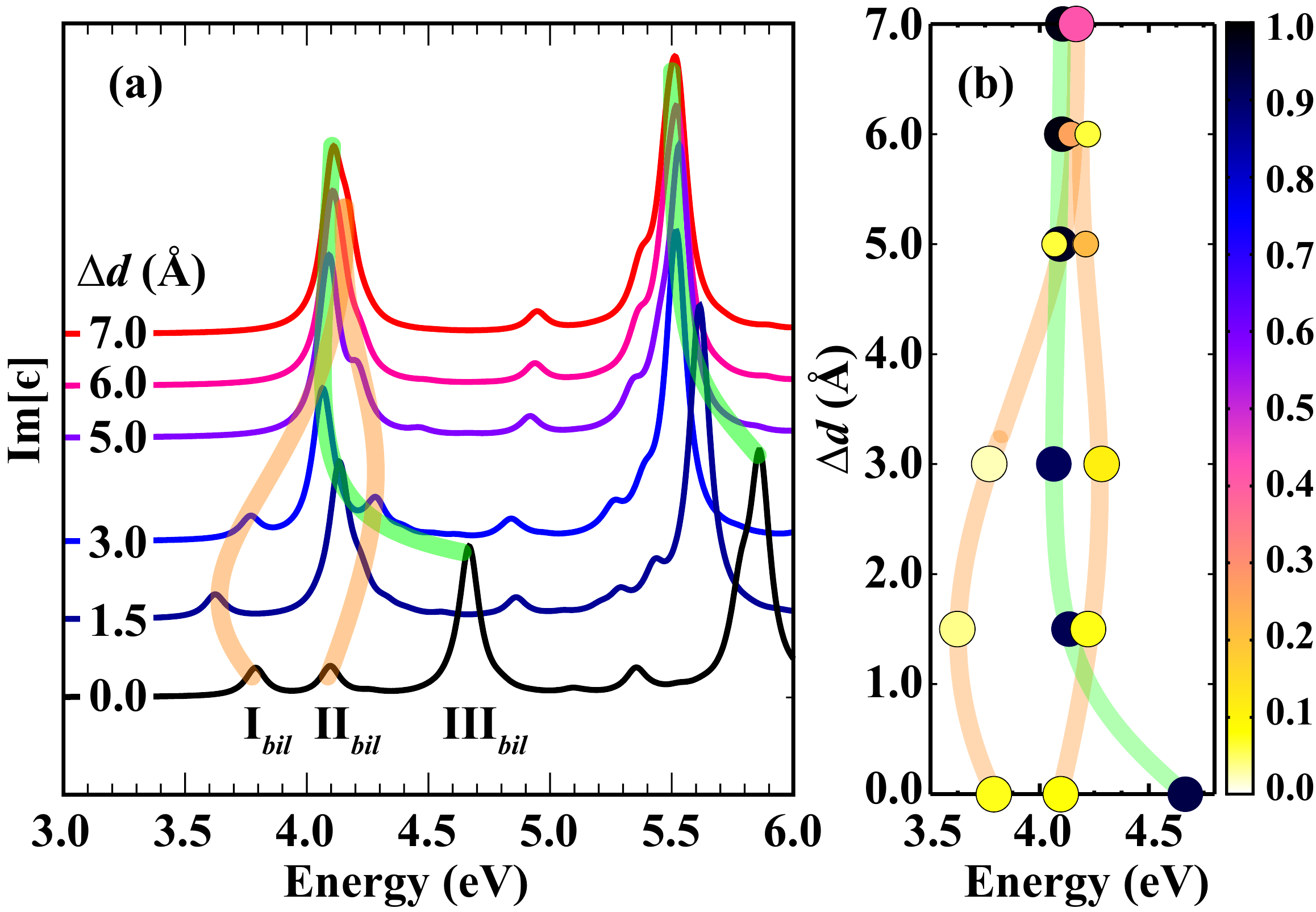}
\caption{\label{dis_osci_fig}
(Color online) (a) The imaginary part of the dielectric functions of the $AB^{\prime}$ stacked hetero-bilayer are shown 
for 
different interlayer distances starting from ground state distance of 2.13 \AA{}. $\Delta d$ refers to interlayer 
distance while the ground state distance is set to 0 and labeled as 0.0 \AA{}. The lowest three exciton peaks are 
labeled as \textbf{I$_{bil}$}, \textbf{II$_{bil}$}, and \textbf{III$_{bil}$}. (b) 
The distance dependent oscillator strengths of lowest three excitons are shown. Color code is given. The position of 
the prominent peaks are illustrated by shaded curves. The length of the out-of-plane lattice vector is fixed for 
different 
interlayer distance. }
\end{figure}

In Fig.~\ref{dis_osci_fig}, we show interlayer-spacing-dependent excitonic properties to investigate the origin and 
evolution of the excitons by tracking their peak positions and corresponding oscillator strengths in the energy 
spectrum.

As shown in Fig.~\ref{dis_osci_fig} (a), starting from the interlayer spacing of 6.0 \AA{} (equilibrium distance is 
set to 0 \AA), the first peak of the isolated Mg(OH)$_{2}$ splits into two different peaks due to the broken symmetry 
of the top and bottom surface charge densities. It is seen that the first peak of the bilayer structure, 
\textbf{I$_{bil}$}, appears at the lower energy level than the first peak of both individual $h$-AlN and Mg(OH)$_{2}$. 
As calculated in the previous section, the VBM and CBM electronic states are mainly formed by the O and N, 
respectively. Therefore, the first peak of the bilayer heterostructure, \textbf{I$_{bil}$}, corresponds to a spatially 
indirect exciton in which the electron localizes at the N of $h$-AlN and the hole is at O of Mg(OH)$_{2}$. As shown in 
Fig.~\ref{dis_osci_fig} (b), the reduction in the oscillator strength of \textbf{I$_{bil}$} (minimum of all) at around 
3.0 \AA{}, reveals that electronic state of Mg(OH)$_{2}$ at the intimate contact surface overlaps with $h$-AlN layer as 
the distance decreases. However, at the optimum interlayer distance, other Mg(OH)$_{2}$-originated exciton state 
evolves into the second peak of the heterostructure labeled as \textbf{II$_{bil}$}. Apparently, the electronic state of 
\textbf{II$_{bil}$} exciton peak originates from the outer surface state of Mg(OH)$_{2}$.

On the other hand, the origin and interlayer-spacing-dependence of \textbf{III$_{bil}$},  differs from those of 
\textbf{I$_{bil}$} and \textbf{II$_{bil}$}. Firstly, in Fig.~\ref{dis_osci_fig}, it is evident that by decreasing 
interlayer spacing the first excitonic peak of $h$-AlN crystal gradually evolves into third state of hetero-bilayer, 
labeled as \textbf{III$_{bil}$}. It is seen from Figs.~\ref{dis_osci_fig} (a) and (b) that the shape and the oscillator 
strength of the \textbf{III$_{bil}$} do not change significantly. Therefore, one can conclude that the characteristic 
properties of the \textbf{III$_{bil}$} are almost same with the first excitonic state of $h$-AlN monolayer crystal. 
When the distance decrease under 4.0 \AA{}, a significant charge distribution appears on the O atom. It appears that 
the screening on the $h$-AlN increases when the Mg(OH)$_{2}$ layers become closer. Therefore as a consequence of 
decrease in exciton binding energy corresponding peak shows a significant blue shift. 

\begin{figure*}[htbp]
\includegraphics[width=17 cm]{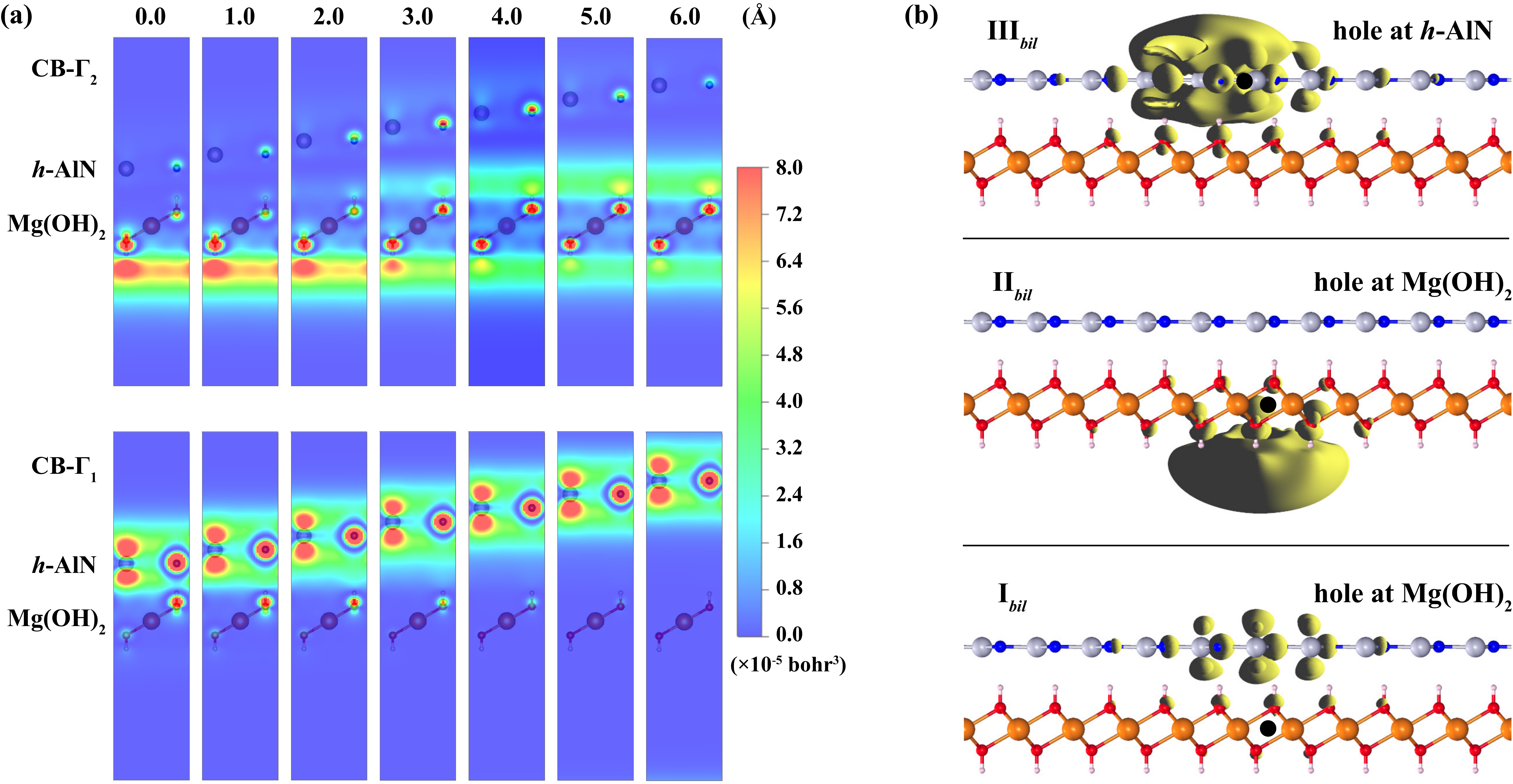}
\caption{\label{chg_dis}
(Color online) (a) Layer-layer distance dependent band decomposed charge densities at the $\Gamma$ point for the lowest 
two bands (CB-$\Gamma_{1}$ and CB-$\Gamma_{2}$) are presented. The the numbers on the upper part of the figure 
are 
distances between the layers in the unit of \AA{}. The optimized layer-layer distance is set to 0.0 \AA{}. (b) shows 
the wave function of the lowest three excitons which are labeled as \textbf{I$_{bil}$}, \textbf{II$_{bil}$}, and 
\textbf{III$_{bil}$}. The yellow regions are the probability of the electron localization 
when the hole is at a specific point. The position of holes
are shown as a black point on the structures for all wave functions. The methodology for the determination of 
excitonic wave functions is given in Appendix \ref{wavefunc}.}
\end{figure*}

For a deeper understanding of nature of excitonic states of the vdW Mg(OH)$_{2}$/$h$-AlN heterostructure we also 
investigate the band decomposed charge densities and exctionic wave functions. In Fig.~\ref{chg_dis} (a), the 
interlayer-spacing-dependent band decomposed charge densities of the first and second 
lowest levels of the CB at $\Gamma$ point are shown. The lowest and second lowest levels are labeled as 
CB-$\Gamma_{1}$ and CB-$\Gamma_{2}$, respectively. The 
changes in the charge densities of these two levels contain remarkable informations about the evolution of the 
electrons involved in the states \textbf{I$_{bil}$}, \textbf{II$_{bil}$}, and \textbf{III$_{bil}$}. 

Firstly, when the distance is 6.0 \AA{}, the charge density shown in CB-$\Gamma_{2}$ part of the Fig.~\ref{chg_dis} (a) 
corresponds to the charge density at the CBM of individual Mg(OH)$_{2}$. It is seen that this density appears mostly on 
O atoms and also at the surfaces of the Mg(OH)$_{2}$. As the interlayer distance decreases, the charge density vanishes 
at the inner surface and increases at the outer surface. On the other hand, the density shown in the CB-$\Gamma_{1}$ for 
the interlayer distance of 6.0 \AA{} corresponds to the charge density at the CBM of individual $h$-AlN. As the distance 
decreases, charge changes negligibly at the vicinity of $h$-AlN but an additional charge density appears on inner O atom 
when the distance is less than 4.0 \AA{}. It appears that when the layers approach each other, the inner surface 
electron of the Mg(OH)$_{2}$ is transfered to the $h$-AlN and the outer surface electron states of the Mg(OH)$_{2}$ 
remains at its original position.

Furthermore, we also calculate the real 
space exciton localizations by using the methodology given in Appendix 
\ref{wavefunc}. The wave functions of the \textbf{I$_{bil}$}, \textbf{II$_{bil}$}, and 
\textbf{III$_{bil}$} are shown in Fig.~\ref{chg_dis} (b) where the yellow regions correspond to electrons and the 
black points 
stand for holes.
As shown in the lower panel, it is evident that the \textbf{I$_{bil}$} is the spatially indirect exciton in which 
the electron and hole localize on the $h$-AlN and Mg(OH)$_{2}$, respectively. 
The \textbf{II$_{bil}$} localizes at the outer side of the Mg(OH)$_{2}$ 
as shown in middle panel of the Fig.~\ref{chg_dis} (b). It seems that the \textbf{II$_{bil}$} is not affected by the 
$h$-AlN. The \textbf{III$_{bil}$} shown in upper panel of the Fig.~\ref{chg_dis} (b) stays mostly on the 
$h$-AlN. A small portion of the \textbf{III$_{bil}$} appears on the hydroxide regions of the Mg(OH)$_{2}$.

\section{Conclusion}\label{conc}

In this study, structural, electronic, and optical properties of the recently synthesized novel 2D materials of $h$-AlN 
and Mg(OH)$_{2}$ and their bilayer heterostructure are investigated by performing DFT calculation. The excitonic states 
are also calculated by solving BSE over the $GW_{0}$ approximation on top of DFT. Our investigation revealed that: (i) 
the individual monolayer of the $h$-AlN and Mg(OH)$_{2}$ can form a vertically stacked hetero-bilayer, (ii) when the 
hetero-bilayer formed, 
although the interlayer interaction is weak novel exciton states appear, (iii) while $h$-AlN-states have weak spacing 
dependence, Mg(OH)$_{2}$ states are strongly affected by the presence of a neighboring layer, (iv) exponential increase 
in screening of Mg(OH)$_{2}$ on $h$-AlN-states was also predicted, (v) while the first exciton peak is a spatially 
indirect one, the second and third exciton states are spatially direct states.

Although the constituents have limited optical activity, the heterostructure shows unexpected optical properties. In 
particular, $h$-AlN-Mg(OH)$_{2}$ hetero-bilayer exhibits spatially indirect excitons which is important for the 
optoelectronic application, especially based on photoexcited electron collecting.

\section{Acknowledgments} 

The calculations were performed at TUBITAK ULAKBIM, High Performance and Grid Computing Center 
(TR-Grid e-Infrastructure). CB and RTS acknowledge the support from TUBITAK Project No 114F397. HS acknowledges 
support from Bilim Akademisi-The Science Academy, Turkey under the BAGEP program. HS acknowledges financial support
from the Scientific and Technological Research Council of
Turkey (TUBITAK) under the project number 116C073. AR and AD acknowledge financial 
support from the European Research Council(ERC-2015-AdG-694097), Spanish grant (FIS2013-46159-C3-1-P), Grupos 
Consolidados (IT578-13), and AFOSR Grant No. FA2386-15-1-0006 AOARD 144088, H2020-NMP-2014 project MOSTOPHOS (GA 
no. 646259) and COST Action MP1306 (EUSpec)

\appendix

\section{Convergence Tests}\label{tests}

Theoretical results on optical properties, especially many-body and excitonic effects, are strongly 
depend on computational parameters such as Brillouin Zone sampling and the vacuum spacing between the adjacent layers. 
The calculation with fine parameters, on the other hand, requires large computational resource and time. Therefore, we 
performed calculations to examine the convergence of the frequency dependent imaginary dielectric function with respect 
to k-point sampling and the vacuum spacing.

\begin{figure}[htbp]
\includegraphics[width=8.5 cm]{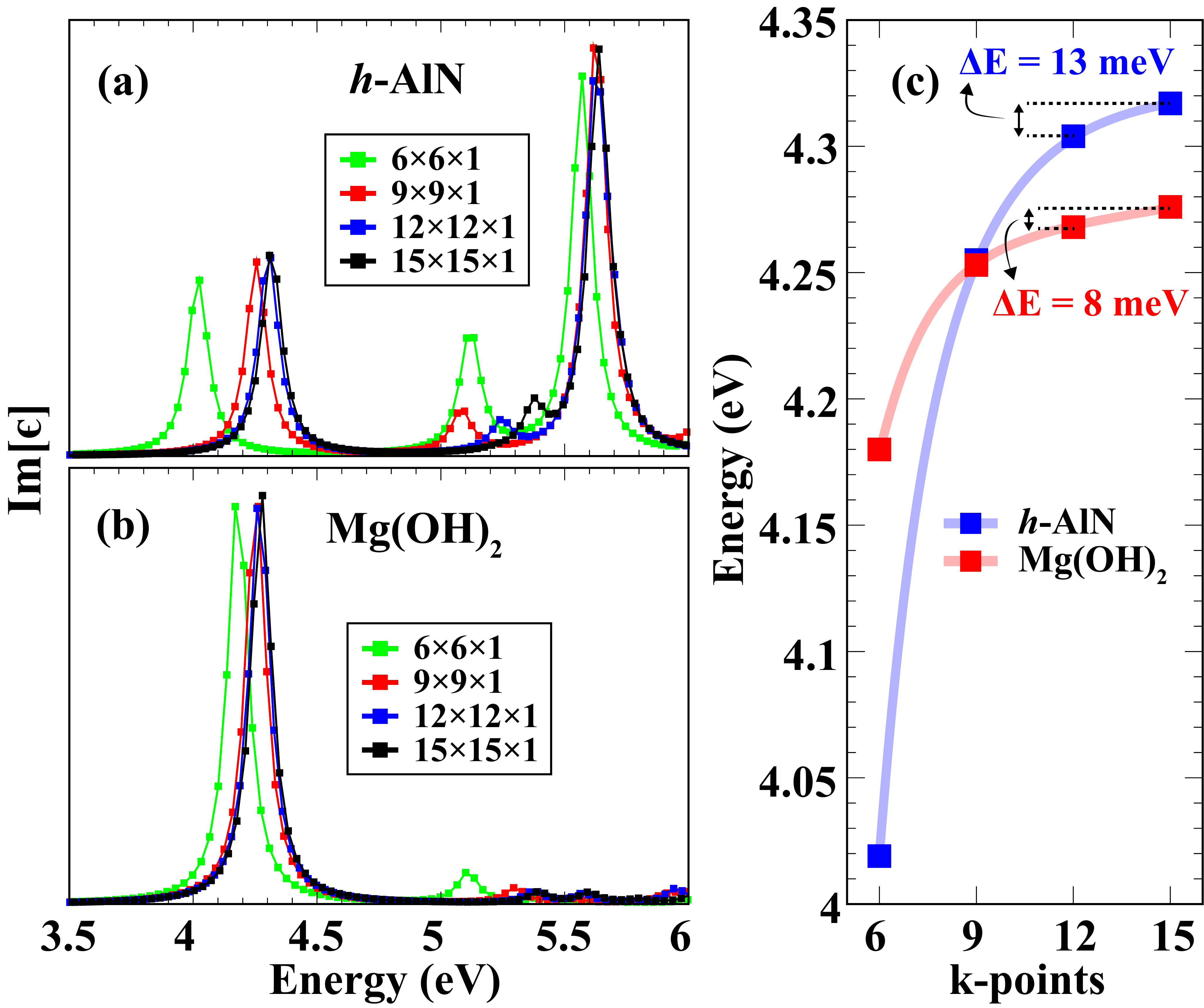}
\caption{\label{k-point}
(Color online) The results of the k-point sampling tests: (a) and (b) The imaginary part of the dielectric functions of 
the monolayer \textit{h}-AlN and Mg(OH)$_{2}$ for different k-point samplings, respectively. (c) The first exciton 
peak positions are shown as the blue and red squares for the monolayer \textit{h}-AlN and 
Mg(OH)$_{2}$, receptively.}
\end{figure}

The imaginary part of the dielectric function for the monolayer 
\textit{h}-AlN and Mg(OH)$_{2}$ with respect to k-point sampling are shown in Fig. \ref{k-point} (a) and (b), 
respectively. When the number of k-points increases, the spectrum of both monolayers are blue-shifted and rapidly 
converges by the 12$\times$12$\times$1 sampling (see Fig. \ref{k-point} (c)). Therefore, using  12$\times$12$\times$1 
for optical properties of monolayer \textit{h}-AlN and Mg(OH)$_{2}$, provides reliable results.

\begin{figure}
\includegraphics[width=8.5 cm]{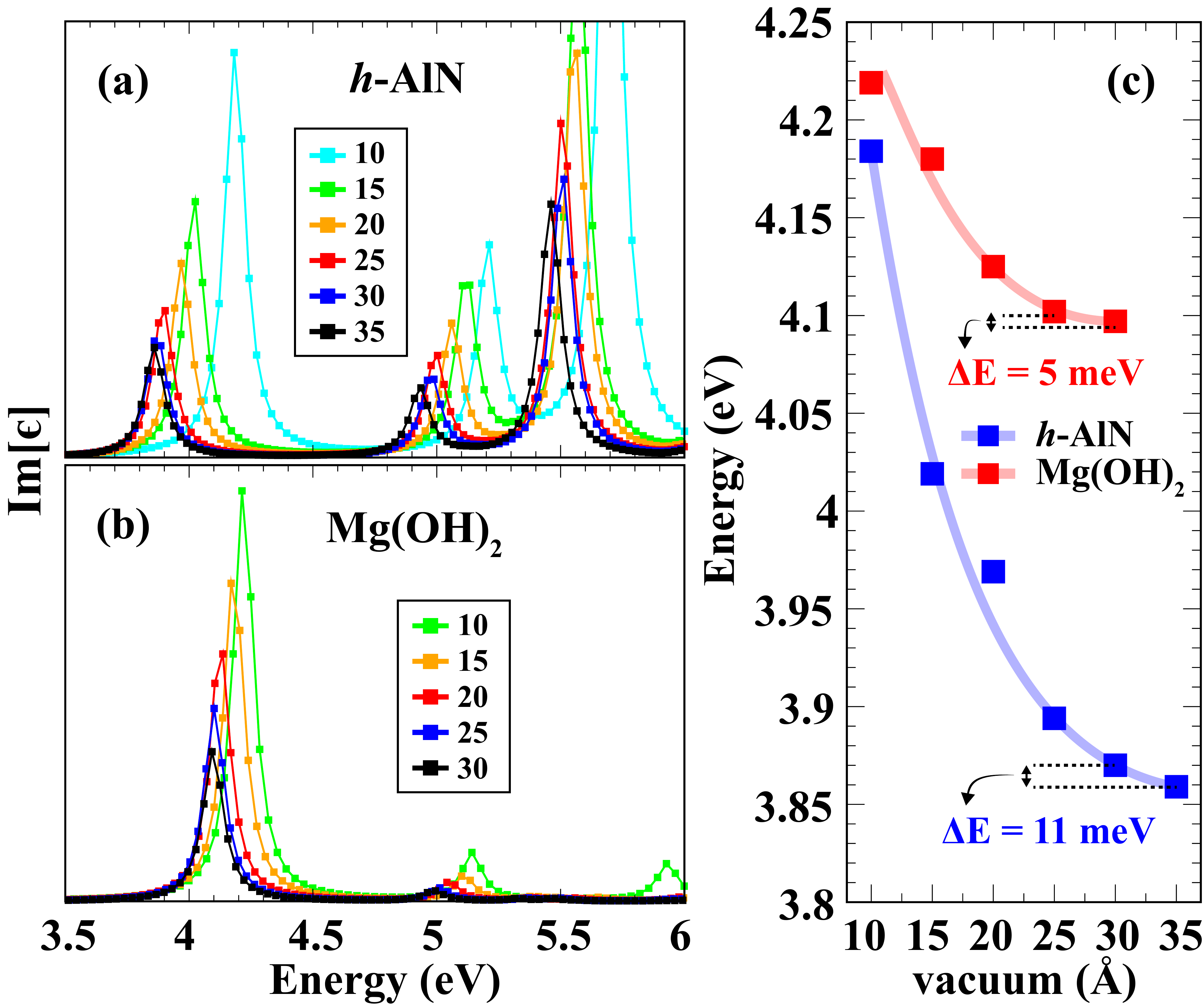}
\caption{\label{vacuum}
(Color online) The results of vacuum spacing tests: (a) and (b) The imaginary part of the dielectric functions 
of the monolayer \textit{h}-AlN and Mg(OH)$_{2}$ for different vacuum spacing values, respectively. (c) The first 
exciton peak positions are shown. The blue and red curves correspond to the monolayer \textit{h}-AlN and Mg(OH)$_{2}$, 
receptively. }
\end{figure}

As shown in Fig. \ref{vacuum} (a) and (b), the effect of the vacuum spacing is also evident that the peak positions of 
both 
monolayers are red-shifted. For the \textit{h}-AlN, the curves of 30 and 35 \AA{} vacuum spacing values are almost the 
same. For Mg(OH)$_{2}$, the curve does not change significantly when the vacuum spacing increases from 25 to 30 \AA{}. 
Moreover, as shown in Fig. \ref{vacuum} (c), the difference between the first exciton energies of 30 and 35 \AA{} 
vacuum spacing values is 11 meV for the monolayer \textit{h}-AlN. The smaller energy difference of 5 meV is obtained 
between 25 and 30 \AA{} vacuum spacing values for the monolayer Mg(OH)$_{2}$. It is clearly seen that, the vacuum 
spacing of 30 and 25 \AA{} are sufficient for the monolayers \textit{h}-AlN and Mg(OH)$_{2}$, respectively. 
Therefore, when we consider the hetero-bilayer, using the vacuum spacing of $\sim$28 \AA{} for optical 
properties provides reliable results. After having convergence tests, the best parameter set of 12x12x1 kpoint and 28 Å 
vacuum spacing are used in our calculations.

\section{Methodology for Determination of Excitonic Wave Functions}\label{wavefunc}

For a deeper understanding of the origin of excitonic states and determination of their wave functions of the vdW 
hetero-bilayer, three lowest exciton wave functions were calculated by using BerkeleyGW package\cite{Deslippe} on top of
QuantumEspresso code.\cite{Giannozzi} The eigenvalues and the eigenfunctions in the independent particle picture were 
obtained by using PBE form of GGA for exchange and correlation. The quasi-particle energies were calculated within the 
GW approximation. We employed an energy cutoff of 160 Ry for the PBE calculations. For the computation of the 
dielectric matrix, we used 1987 conduction bands and G-vectors with energy up to 17 Ry, whereas the self-energy 
operator was computed using 1987 conduction bands and a G-vector cutoff of 17 and 160 Ry for the screened and bare 
Coulomb matrices, respectively. To sample the BZ we employed a $6\times6\times1$ k-point grid for the PBE and GW 
calculations, whereas for the solution of the BSE, we used a $18\times18\times3$ k-point grid with linearly 
interpolated GW quasi-particles energies. The 8 lowest conduction bands and 8 topmost valence bands were included to 
solve the BSE. After obtaining the excitonic states, the probability of localization of the electron is calculated for 
each excitonic level when the hole is fixed to a specific point. The result are given in Fig. \ref{chg_dis} (b).


\begin{thebibliography}{99}

\bibitem{Novoselov2} K. S. Novoselov, A. K. Geim, S. V. Morozov, D. Jiang, Y. Zhang, S. V. Dubonos, 
I. V. Grigorieva, and A. A. Firsov, Science \textbf{306}, 666 (2004).

\bibitem{Novoselov1}K. S. Novoselov, D. Jiang, F. Schedin, T. J. Booth, V. V. Khotkevich, S. V. 
Morozov, and A. K. Geim, Proc. Natl. Acad. Sci. U.S.A. \textbf{102}, 10451 (2005).

\bibitem{Butler}  S. Z. Butler, S. M. Hollen, L. Cao,  Y. Cui, J. A. Gupta, H. R. Gutié rrez, T. F. 
Heinz, S. S. Hong, J. Huang, A. F. Ismach, E. Johnston-Halperin, M. Kuno, V. V. Plashnitsa, R. D. 
Robinson, R. S. Ruoff, S. Salahuddin, J. Shan, L. Shi, M. G. Spencer, M. Terrones, W. Windl, and J. 
E. Goldberger, ACS Nano \textbf{7}, 2898 (2013).

\bibitem{Chhowalla} M. Chhowalla, H. S. Shin, G. Eda, L-J Li, K. P. Loh, H Zhang, Nat. Chem. 
\textbf{5}, 263 (2013).

\bibitem{Cahangirov} S. Cahangirov, M. Topsakal, E. Akturk, H. Sahin, and S. Ciraci, Phys. Rev. 
Lett. \textbf{102}, 236804 (2009). 

\bibitem{Kara} A. Kara, H. Enriquez, A. P. Seitsonen, L. C. L. Y. Voon, S. Vizzini, B. Aufray, and 
Hamid Oughaddou, Surf. Science Report. \textbf{67}, 1 (2012).
 
\bibitem{Guzman} G. G. Guzman-Verri and L. C. Lew Yan Voon, Phys.Rev.B 76, 075131 (2007).

\bibitem{Bechstedt} F. Bechstedt, L. Matthes, P. Gori, and O. Pulci, Appl. Phys. Lett. 100, 261906 
(2012).

\bibitem{Gordon} R. A. Gordon, D. Yang, E. D. Crozier, D. T. Jiang, and R. F. Frindt, Phys. Rev. B 
\textbf{65}, 125407 (2002).

\bibitem{Coleman} J. N. Coleman, M. Lotya, A. O'Neill, S. D. Bergin, P. J. King, U. Khan, K. Young, 
A. Gaucher, S. De, R. J. Smith, I. V. Shvets, S. K. Arora, J. J. Boland, J. J. Wang, J. F. Donegan, 
J. C. Grunlan, G. Moriarty, A. Shmeliov, R. J. Nicholls, J. M. Perkins, E. M. Grieveson, K. 
Theuwissen, D. W. McComb, P. D. Nellist, and V. Nicolosi, Science \textbf{331}, 568 (2011). 

\bibitem{Wang1} Q. H. Wang, K. Kalantar-Zadeh, A. Kis, J. N. Coleman, and M. S. Strano, Nature 
Nanotech. \textbf{7}, 699 (2012).

\bibitem{Ross} J. S. Ross, P. Klement, A. M. Jones, N. J. Ghimire, J. Yan, D. G. Mandrus, T. 
Taniguchi, K. Watanabe, K. Kitamura, W. Yao, D. H. Cobden, and X. Xu, Nature Nanotech. \textbf{9} 
268 (2014).

\bibitem{Sahin2} H. Sahin, S. Tongay, S. Horzum, W. Fan, J. Zhou, J. Li, J. Wu, and F. M. Peeters, 
Phys. Rev. B \textbf{87}, 165409 (2013).

\bibitem{Tongay} S. Tongay, H. Sahin, C. Ko, A. Luce, W. Fan, K. Liu, J. Zhou, Y.-S. Huang, C.-H. 
Ho, J. Yan, D. F. Ogletree, S. Aloni, J. Ji, S. Li, J. Li, F. M. Peeters, and J. Wu, Nat. Comm. 
\textbf{5} 3252 (2014). 

\bibitem{Horzum} S. Horzum, D. Cakir, J. Suh, S. Tongay, Y.-S. Huang, C.-H. Ho, J. Wu, H. Sahin, 
and F. M. Peeters, Phys. Rev. B \textbf{89}, 155433 (2014).

\bibitem{Chen3} B. Chen, H. Sahin, A. Suslu, L. Ding, M. I. Bertoni, F. M. Peeters, and S. Tongay, 
ACS Nano \textbf{9}, 5326 (2015).

\bibitem{Sahin3} H. Sahin, S. Cahangirov, M. Topsakal, E. Bekaroglu, E. Akturk, R. T. Senger, and S. 
Ciraci, Phys. Rev. B \textbf{80}, 155453 (2009).

\bibitem{Wang2} Q. Wang, Q. Sun, P. Jena, and Y. Kawazoe, ACS Nano \textbf{3}, 621 (2009).

\bibitem{Kim} K. K Kim, A. Hsu, X. Jia, S. M. Kim, Y. Shi, M. Hofmann, D. Nezich, J. F. 
Rodriguez-Nieva, M. Dresselhaus, T. Palacios, and J. Kong, Nano Lett. \textbf{12}, 161 (2012).

\bibitem{Tsipas}  P. Tsipas, S. Kassavetis, D. Tsoutsou, E. Xenogiannopoulou, E. Golias, S. A. 
Giamini, C. Grazianetti, D. Chiappe, A. Molle, M. Fanciulli, and A. Dimoulas, Appl. Phys. Lett. 
\textbf{103}, 251605 (2013).

\bibitem{Bacaksiz} C. Bacaksiz, H. Sahin, H. D. Ozaydin, S. Horzum, R. T. Senger, and F. M. 
Peeters, Phys. Rev. B \textbf{91}, 085430 (2015).


\bibitem{Geim} A. Geim and I. Grigorieva, Nature (London) \textbf{499}, 419 (2013).


\bibitem{Britnell} L. Britnell, R. Gorbachev, R. Jalil, B. Belle, F. Schedin, A. Mishchenko, T. 
Georgiou, M. Katsnelson, L. Eaves, and S. Morozov, Science 335, 947 (2012).

\bibitem{Fang} H. Fang, C. Battaglia, C. Carraro, S. Nemsak, B. Ozdol, J. S. Kang, H. A. Bechtel, 
S. B. Desai, F. Kronast, and A. A. Unal, Proc. Natl. Acad. Sci. USA 111, 6198 (2014).

\bibitem{Lee2} C.-H. Lee, G.-H. Lee, A. M. van der Zande, W. Chen, Y. Li, M. Han, X. Cui, G. Arefe, 
C. Nuckolls, and T. F. Heinz, Nat. Nanotechnol. 9, 676 (2014).

\bibitem{Hunt} B. Hunt, J. Sanchez-Yamagishi, A. Young, M. Yankowitz, B. J. LeRoy, K. Watanabe, T. 
Taniguchi, P. Moon, M. Koshino, and P. Jarillo-Herrero, Science 340, 1427 (2013).

\bibitem{Hong} X. Hong, J. Kim, S.-F. Shi, Y. Zhang, C. Jin, Y. Sun, S. Tongay, J. Wu, Y. Zhang, 
and F. Wang, Nat. Nanotechnol. \textbf{9}, 682 (2014).

\bibitem{Ferrari} A. C. Ferrari et al., Nanoscale \textbf{7}, 4598 (2015).

\bibitem{Tan} C. L. Tan and H. Zhang, Chem. Soc. Rev. \textbf{44}, 2713 (2015).


\bibitem{Almeida} E. F. de Almeida Junior, F. de Brito Mota, C. M. C. de Castilho, A. 
Kakanakova-Georgieva, and G. K. Gueorguiev, Eur. Phys. J. B. \textbf{85}, 48 (2012).

\bibitem{Shi} C. Shi, H. Qin, Y. Zhang, J. Hu, and L. Ju, J. Appl. Phys. \textbf{115}, 053907 
(2014).

\bibitem{Zheng} F. L. Zheng, J. M. Zhang, Y. Zhang, and V. Ji, Physica B \textbf{405}, 3775 (2010).

\bibitem{Kecik} D. Kecik, C. Bacaksiz, R. T. Senger, and E. Durgun, Phys. Rev. B \textbf{92}, 
165408 (2015).

\bibitem{Utamapanya} S. Utamapanya, K. J. Klabunde, and J. R. Schlup, Chem. Mater. \textbf{3}, 175 
(1991).

\bibitem{Ding} Y. Ding, G. Zhang, H. Wu, B. Hai, L. Wang, and Y. Qian, Chem. Mater. \textbf{13}, 
435 (2001).

\bibitem{Sideris} P. J. Sideris, U. G. Nielsen, Z. Gan, and C. P. Grey, Science \textbf{321}, 113 
(2008).


\bibitem{Murakami} T. Murakami, T. Honjo, and T. Kuji, Mater. Trans. \textbf{52}, 1689 (2011).

\bibitem{Aierken} Y. Aierken, H. Sahin, F. Iyikanat, S. Horzum, A. Suslu, B. Chen, R. T. Senger, S. 
Tongay, and F. M. Peeters, Phys. Rev. B \textbf{91}, 245413 (2015).

\bibitem{Torun2} E. Torun, H. Sahin, and F. M. Peeters, Phys. Rev. B \textbf{93}, 075111 (2016).

\bibitem{Tsukanov} A. A. Tsukanov and S. G. Psakhie, Sci. Rep. \textbf{6}, 19986 (2016).


\bibitem{Suslu} A. Suslu, K. Wu, H. Sahin, B. Chen, S. Yang, H. Cai, T. Aoki, S. Horzum, J. Kang, 
F. M. Peeters, and S. Tongay, Sci. Rep. \textbf{6}, 20525 (2016).

\bibitem{Yagmurcukardes} M. Yagmurcukardes, E. Torun, R. T. Senger, F. M. Peeters, and H. Sahin,
Phys. Rev. B \textbf{94}, 195403 (2016).

\bibitem{vasp1} G. Kresse and J. Hafner, Phys. Rev. B \textbf{47}, 558 (1993).

\bibitem{vasp2} G. Kresse and J. Furthmuller, Phys. Rev. B \textbf{54}, 11169 (1996).

\bibitem{vasp3} G. Kresse and D. Joubert, Phys. Rev. B \textbf{59}, 1758 (1999).


\bibitem{GGA-PBE} J. P. Perdew, K. Burke, and M. Ernzerhof, Phys. Rev. Lett. \textbf{77}, 3865 
(1996).



\bibitem{vdW1} S. J. Grimme, Comput. Chem. \textbf{27}, 1787 (2006).

\bibitem{vdW2} T. Bucko, J. Hafner, S. Lebegue, and J. G. Angyan,  J. Phys.Chem. A \textbf{114}, 
11814 (2010).



\bibitem{Bader1} R. F. W. Bader, Atoms in Molecules – A Quantum Theory(Oxford University Press, 
Oxford, UK, 1990).

\bibitem{bse} H. Bethe and E. Salpeter, Phys. Rev. \textbf{84}, 1232 (1951).

\bibitem{Hanke} W. Hanke and L. J. Sham, Phys. Rev. B \textbf{21}, 4656 (1980).






\bibitem{Mak} K. F. Mak, C. Lee, J. Hone, J. Shan, and T. F. Heinz, Phys. Rev. Lett. \textbf{105}, 
136805 (2010).

 \bibitem{Wirtz} L. Wirtz, A. Marini, and A. Rubio Phys. 
Rev. Lett. \textbf{96}, 126104 (2006).

\bibitem{Pishtshev} A. Pishtshev, S. Zh. Karazhanov, and M. Klopov, Solid State Commun. 
\textbf{193}, 11 (2014). 

 \bibitem{Cudazzo} P. Cudazzo, L. Sponza, C. Giorgetti, L. Reining, F. Sottile, and M. Gatti, Phys. 
Rev. Lett. \textbf{116}, 066803 (2016).

\bibitem{Deslippe} J. Deslippe, G. Samsonidzea, D. A. Strubbea, M. Jaina, and M. L. Cohen, Comput. Phys. Commun. 
\textbf{183}, 1269 (2012).

\bibitem{Giannozzi} P. Giannozzi, S. Baroni, N. Bonini, M. Calandra, R. Car, C. Cavazzoni, D. Ceresoli, G. L 
Chiarotti, M. Cococcioni, I. Dabo, A. D. Corso, S. de Gironcoli, S. Fabris, G. Fratesi, R. Gebauer, U. Gerstmann1, C. 
Gougoussis, A. Kokalj, M. Lazzeri5, L. Martin-Samos, N. Marzari, F. Mauri, R. Mazzarello, S. Paolini, A. Pasquarello, 
L. Paulatto, C. Sbraccia1, S. Scandolo, G. Sclauzero, A. P Seitsonen, A. Smogunov, P. Umari, and R. M Wentzcovitch, J. 
Phys. Condens. Matter \textbf{21}, 395502 (2009).


\end{thebibliography}
\end{document}